\documentclass[twocolumn]{aastex631}
\usepackage{CJK}
\usepackage{amsmath}

\begin{document}
\begin{CJK*}{UTF8}{gbsn}

\title{The $\sim$50-Myr-Old TOI-942c is Likely on an Aligned, Coplanar Orbit and Losing Mass}

\author[0000-0003-3860-6297]{Huan-Yu Teng (滕环宇)}
\correspondingauthor{Huan-Yu Teng}
\email{hyteng@bao.ac.cn} 
\affiliation{CAS Key Laboratory of Optical Astronomy, National Astronomical Observatories, Chinese Academy of Sciences, Beijing 100101, China}
\affiliation{Korea Astronomy and Space Science Institute
776 Daedeok-daero, Yuseong-gu, Daejeon 34055, Republic of Korea}

\author[0000-0002-8958-0683]{Fei Dai (戴飞)}
\affiliation{Institute for Astronomy, University of Hawai`i, 2680 Woodlawn Drive, Honolulu, HI 96822, USA}
\affiliation{Division of Geological and Planetary Sciences,
1200 E California Blvd, Pasadena, CA, 91125, USA}
\affiliation{Department of Astronomy, California Institute of Technology, Pasadena, CA 91125, USA}

\author[0000-0001-8638-0320]{Andrew W. Howard}
\affiliation{Department of Astronomy, California Institute of Technology, Pasadena, CA 91125, USA}

\author[0000-0002-0531-1073]{Howard Isaacson}
\affiliation{{Department of Astronomy,  University of California Berkeley, Berkeley CA 94720, USA}}
\affiliation{Centre for Astrophysics, University of Southern Queensland, Toowoomba, QLD, Australia}

\author[0000-0003-3856-3143]{Ryan A. Rubenzahl}
\altaffiliation{NSF Graduate Research Fellow}
\affiliation{Department of Astronomy, California Institute of Technology, Pasadena, CA 91125, USA}

\author[0000-0002-9751-2664]{Isabel Angelo}
\affil{Department of Physics \& Astronomy, University of California Los Angeles, Los Angeles, CA 90095, USA}

\author[0000-0001-7047-8681]{Alex S. Polanski} 
\affil{Department of Physics and Astronomy, University of Kansas, Lawrence, KS 66045, USA}




\begin{abstract}
We report the observation of the transiting planet TOI-942c, a Neptunian planet orbiting a young K-type star approximately 50 Myr years old. Using Keck/HIRES, we observed a partial transit of the planet and detected an associated radial velocity anomaly. By modeling the Rossiter-McLaughlin (RM) effect, we derived a sky-projected obliquity of $\left|\lambda\right|=24^{+14}_{-14}$\,degrees, indicating TOI-942c is in a prograde and likely aligned orbit. Upon incorporation of the star's inclination and the planet's orbital inclination, we determined a true obliquity for TOI-942c of $\psi< 43$\,degrees at 84\% confidence, while dynamic analysis strongly suggests TOI-942c is aligned with stellar spin and coplanar with the inner planet. Furthermore, TOI-942c is also a suitable target for studying atmospheric loss of young Neptunian planets that are likely still contracting from the heat of formation. We observed a blueshifted excess absorption in the H-alpha line at 6564.7 \AA, potentially indicating atmospheric loss due to photoevaporation. However, due to the lack of pre-ingress data, additional observations are needed to confirm this measurement.

\end{abstract}



\section{Introduction}\label{sec:intro}
Stellar obliquity, a probe to planet migration history, is the angle between the rotation axis of the host star and the normal of the orbital plane of the planet. The sky-projected stellar obliquity $\lambda$ can be measured by the Rossiter-McLaughlin (RM) effect \citep{Rossiter1924,McLaughlin1924}, which reveals the distortion of stellar spectral lines during the transits due to the  partial occultation of the rotating stellar surface. This phenomenon can manifest as a pattern of anomalous radial velocity (RV) variation or line shape variation during a planetary transit (e.g., \citealt{Ohta2005, Fabrycky2009}). The true obliquity $\psi$ can subsequently be estimated by combining stellar inclination on the plane of the sky and the orbital inclination of the planet.

A low stellar obliquity may suggest a smooth planetary migration history in the disk, while spin-orbit misalignment is typically interpreted as indicative of a dynamically hot formation or evolutionary history. Theories on dynamically hot planetary migration span various timescales and mechanisms, including primordial disk misalignment ($\lesssim 3$ Myr; e.g., \citealt{Lai2011, Batygin2012}), nodal precession (a few Myr, e.g., \citealt{Yee2018}), Kozai-Lidov mechanism ($10^{4}-10^{8}$ yr; e.g., \citealt{Fabrycky2007}), secular chaos between planets ($10^{7}-10^{8}$ yr; e.g., \citealt{Wu2011}). These mechanisms offer differing predictions regarding obliquity. Notably, these orbit-tilting excitations can occur during the early stages of planetary systems, within a few hundred Myr. Therefore, a collection of obliquity measurements from young systems can help distinguish between these mechanisms.

In addition, the existence of a planetary atmosphere can serve as another probe for investigating planet formation and migration. Planetary atmosphere loss can lead to the reductions in planetary radius and mass (e.g., \citealt{Dong2018}), which may result from atmospheric photoevaporation, happening in the system's first few Myr due to X-ray/ultraviolet (XUV) irradiation from the young star (e.g., \citealt{Owen2017,Owen2018,Ginzburg2018}). 

TOI-942 is a young K-type star ($53_{-21}^{+22}$ Myr; \citealt{Wirth2021}) hosting a pair of transiting hot Neptunes on 4 and 10-day orbits \citep{Zhou2021, Carleo2021} discovered by Transiting Exoplanet Survey Satellite (TESS; \citealt{Ricker2014}). The two planets both have radii larger than 4 $R_{\oplus}$, and their host star has a projected rotational velocity $v\sin i_{\star}$ of $\sim15\,\rm{km}\,\rm{s}^{-1}$. With these properties, TOI-942 offers an opportunity for measuring the stellar obliquity for both Neptunian planets in a young system and constraining the mutual inclination in 3D, as well as studying atmosphere of young Neptunian planets.

The rest of the letter is organized as follows. 
In Section \ref{sec:obs}, we describe our spectroscopic observation of TOI-942c with Keck/HIRES. 
In Section \ref{sec:star}, we describe our constraints on stellar parameters of the host star.
In Section \ref{sec:rm}, we explain our joint analysis of RM effect with TESS light curve and present the stellar obliquity results.
Finally, in Section \ref{sec:discussion}, we discuss our results and implications from this work.

\section{Spectroscopic Observations}\label{sec:obs}
We observed the RM effect of TOI-942c during the transit event on 2021 November 25th (UTC) with High Resolution Echelle Spectrometer (HIRES; \citealt{Vogt1994}) on the 10m Keck I Telescope at Maunakea. The spectrograph has a resolution of $\sim$60000 and covers a wavelength range of 3643--7990 \AA. An iodine cell was equipped in the optical path to the spectrograph, which provides numerous iodine absorption lines in the range of 5000 -- 6000 \AA\, as a reference for precise RV measurements. The spectrograph also enables tracing of the H$\alpha$ line profile variation at 6564 \AA. 

\begin{figure}
\centering
\includegraphics[width=0.99\linewidth]{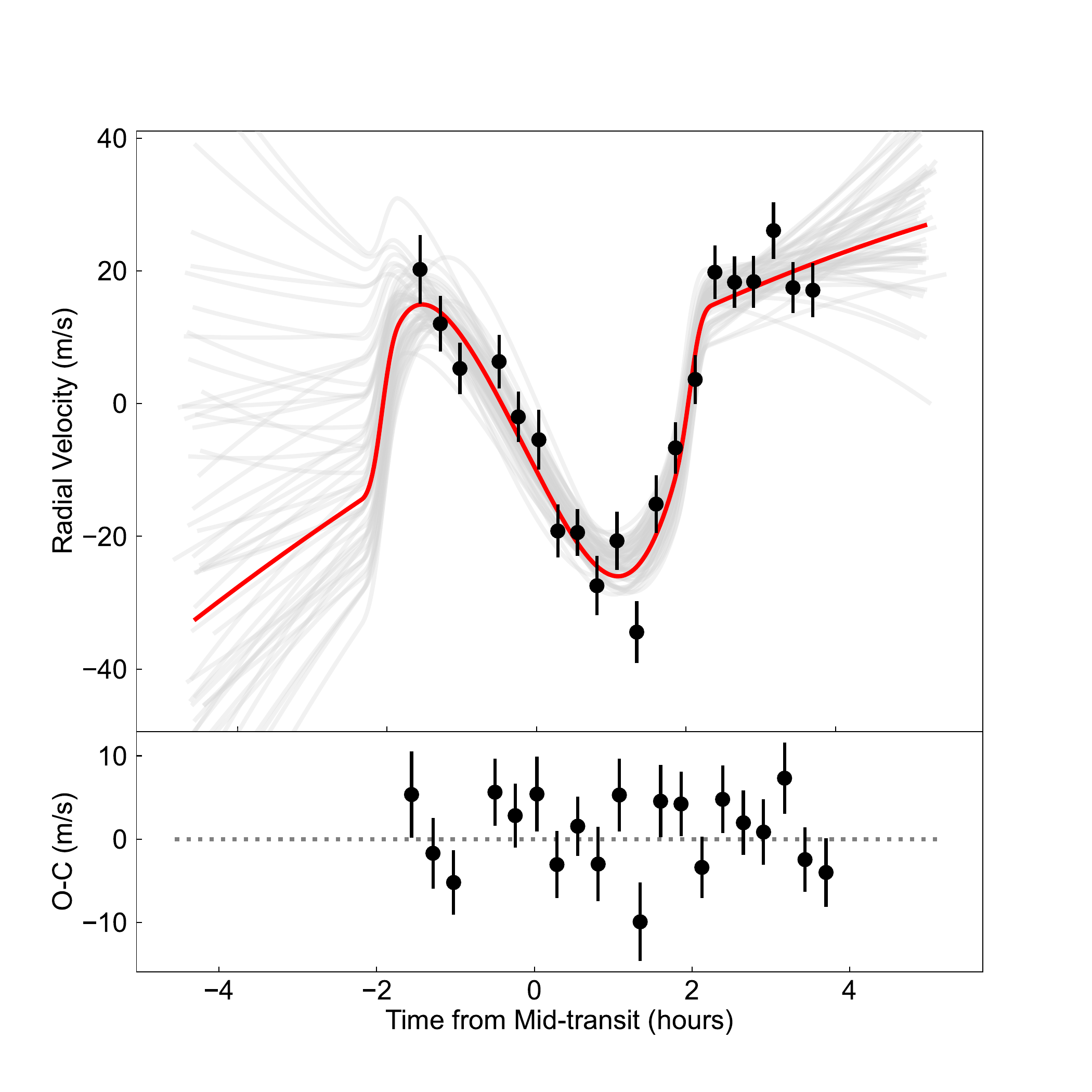}
\caption{The radial velocities (RVs) during the transit event of TOI-942c on 2021 November 25th (UTC), measured by Keck/HIRES spectra. In the upper subplot, the solid red line is the best-fit model ($\left|\lambda\right| = 24^{+14\circ}_{-14}$), incorporating Rossiter-McLaughlin effect and RV long-term trend. The solid gray lines are the solutions randomly selected from the Markov Chain Monte Carlo sampling. In the lower panel, the residuals are given by subtracting the best-fit model. \label{fig:rm}}
\end{figure}

Our observations were scheduled based on the ephemeris from \citet{Zhou2021}. However, due to the underestimation of transit duration, our observation started about 0.7 hours after the planet's ingress and ended $\sim$1.5 hours after the egress and did not encompass the entire transit. 
We set an exposure time of 900 s and reached a signal-to-noise ratio (S/N) of $\sim$80--100 near 5500 \AA. In total, we obtained 21 usable spectra with 15 of them during the transit event. Another high S/N and iodine-free spectrum was obtained separately and used to generate the template spectrum to extract RV values from iodine superimposed spectra.
Details of RV extraction with forward-modeling Doppler code can be found in \citet{Howard2010}, and the RV data shown in Figure \ref{fig:rm} is listed in Table \ref{tab:data}.

\section{Stellar Parameters}\label{sec:star}
\begin{deluxetable}{lccc}
\tablecaption{Stellar Parameters of TOI-942} \label{tab:stellar_para}
\tablehead{
\colhead{Parameters (Unit)} &  \colhead{Value and Uncertainty} & \colhead{Reference}}
\startdata
TIC ID & 146520535 & A\\
R.A. & 05:06:35.91 & A\\
Dec. & -20:14:44.21 & A\\
V (mag)  & 11.982 $\pm$ 0.026& A\\
K (mag)  & 9.639 $\pm$ 0.023& A\\
$T_{\text{eff}} ~(K)$ & $5097\pm100$ & B \\
$\log~g~(\text{dex})$ &$4.47 \pm 0.10$& B \\
$[\text{Fe/H}]~(\text{dex})$ &$0.17 \pm 0.06$& B \\
$v\sin\,i_{\star}$ ($\rm{km}\,\rm{s}^{-1}$) &$15.0 \pm 1.0$ & B$^{\dagger}$ \\
$v\sin\,i_{\star}$ ($\rm{km}\,\rm{s}^{-1}$) &$14.26 \pm 0.50$ & C$^{\dagger}$ \\
$v\sin\,i_{\star}$ ($\rm{km}\,\rm{s}^{-1}$) &$13.8 \pm 0.5$ & D$^{\dagger}$ \\
$R_{\star} ~(R_{\odot})$ &$0.84\pm0.04$& B \\
$M_{\star} ~(M_{\odot})$ &$0.86\pm0.03$& B \\
$P_{\rm{rot}}$ (days) & $3.40 \pm 0.37$ & E \\
$P_{\rm{rot}}$ (days) & $3.39 \pm 0.01$ & D \\
Age (Myr) & $50^{+30}_{-20}$ & E \\ 
Age (Myr) & $53^{+22}_{-21}$ & C \\
\enddata
\tablecomments{A: TICv8; B: This work; C: \citet{Wirth2021}; D:\citet{Carleo2021} E: \citet{Zhou2021};  $^{\dagger}$These three $v\sin\,i_{\star}$ results are derived from spectroscopy.}
\end{deluxetable}
We used iodine-free stellar spectrum obtained from Keck/HIRES and employed the \texttt{SpecMatch-Syn} pipeline\footnote{\url{https://github.com/petigura/specmatch-syn}} \citep{Petigura2017} to determine the spectroscopic parameters of TOI-942, including stellar effective temperature $T_{\rm{eff}}$, surface gravity $\log g$, metallicity $\rm{[Fe/H]}$ as well as the projected rotational velocity $v\sin i_{\star}$. To summarize the methodology, the pipeline models the observed spectra with a synthetic model spectra, by an integration over a precomputed grid \citep{Coelho2005} within the discrete parameter space established on ($T_{\rm{eff}}$, $\log g$, $\rm{[Fe/H]}$, $v\sin i_{\star}$).

We further estimated stellar parameters following the procedure as described in \citet{Fulton2018} with the \texttt{Isoclassify} package \citep{Huber2017}. In principle, we first applied the Stefan–Boltzmann law to derive the radius of the star from posterior probability distributions with inputs including stellar effective temperature, the parallax measurement from Gaia \citep{Gaia2021} and the $K$-band magnitude (which is less affected by extinction). Then we determined various stellar parameters (e.g., stellar mass) from posterior distribution by integrating over the the MESA Isochrones \& Stellar Tracks \citep[MIST,][]{Choi2016} with spectroscopic parameters and the parallax and their priors. 

Consequently, we obtained $R_{\star} = 0.84 \pm 0.04\, R_{\odot}$ for the stellar radius and $M_{\star} = 0.86 \pm 0.03\, M_{\odot}$ for the stellar mass. These results agree well with $R_{\star} = 0.893_{-0.053}^{+0.071}\, R_{\odot}$ and $M_{\star} = 0.880 \pm 0.040\, M_{\odot}$ from \citet{Carleo2021} and $R_{\star} = 0.894_{-0.052}^{+0.056}\, R_{\odot}$ and $M_{\star} = 0.8220_{-0.0064}^{+0.0079}\, M_{\odot}$ from \citet{Wirth2021}, mostly within $\sim$1$\sigma$. Notably, we emphasize the difference in projected rotational velocity, although our result of $15.0 \pm 1.0\, \rm{km}\,\rm{s}^{-1}$ agrees well with $14.26 \pm 0.50\, \rm{km}\, \rm{s}^{-1}$ from \citet{Wirth2021} and $13.8 \pm 0.5\, \rm{km}\, \rm{s}^{-1}$ from \citet{Carleo2021} within $\sim$1$\sigma$. In addition, projected rotational velocity can be estimated from rotational period and stellar radius. Using the stellar radius derived in this work and rotational period of $3.40\pm 0.37$ days from \citet{Zhou2021}, we obtained a maximum velocity of $12.5 \pm 2.0\, \rm{km}\,\rm{s}^{-1}$. This is consistent with the three spectroscopic results within $\sim$1$\sigma$, but suggesting that the spectroscopic results are slightly overestimated and the stellar inclination $i_{\star}$ should be close to $90^{\circ}$. We then calculated stellar inclination using Bayesian method described in \citet{Masuda2020} and obtained a maximum posterior at $89.9^{\circ}$ regardless of which spectroscopic rotational velocity value was used. Importantly, the projected rotational velocity is a key parameter and can be constrained by RM effect modeling. Different priors from spectroscopic measurements produce variations in obliquity constraints. Key stellar parameters are listed in Table \ref{tab:stellar_para}.

\section{Joint Light Curve and RM Analysis}\label{sec:rm}
\begin{deluxetable*}{lcccc}
\tablecaption{Stellar and Transit Parameters of TOI-942c} \label{tab:planet_para}
\tablehead{
\colhead{Parameters (Unit)} & & \colhead{Values and Uncertainties}  & & \colhead{Priors}}
\startdata
& Model 1  & Model 2 & \textbf{Model 3}\\
$v\sin i_{\star}$ Prior ($\rm{km}\,\rm{s}^{-1}$) & None & $\mathcal{N}[15.0, 1.0]$ & $\mathcal{N}[14.26, 0.50]$ &  \\
\hline
$\ln \rho$ ($\rho_{\odot}$) & $-0.53_{-0.29}^{+0.26}$ & $-0.43_{0.26}^{+0.19}$ & $-0.45_{-0.26}^{+0.18}$ & $\mathcal{U}[-1, 2]$ \\
$P_{\rm orb}$ (days)  & $10.15609_{-0.00015}^{+0.00011}$ & $10.15605_{-0.00013}^{+0.00012}$ & $10.15607_{-0.00012}^{+0.00012}$ & $\mathcal{U}[10.1555, 10.1565]$ \\
$t_0$ (BJD-2457000) & $1447.0708_{-0.0088}^{+0.0112}$ & $1447.0736_{-0.0093}^{-0.0101}$ & $1447.0725_{-0.0088}^{-0.0090}$ & $\mathcal{U}[1447.04, 1447.12]$ \\
$R_{\rm{p}}/R_{\star}$  & $0.0530^{+0.0014}_{-0.0012}$& $0.0528_{-0.0011}^{+0.0013}$ & $0.0528_{-0.0011}^{+0.0013}$ & $\mathcal{U}[0.02, 0.08]$  \\
$\cos\,i_{\rm{orb}}$ & $0.027_{-0.016}^{+0.012}$ & $0.023_{-0.014}^{+0.012}$ & $0.024_{-0.011}^{+0.013}$& $\mathcal{U}[0, 1]$ \\
$\sqrt{v\sin i_{\star}}\cos\lambda$ & $3.28_{-0.65}^{+0.45}$ & $3.49_{-0.52}^{+0.30}$ & $3.45_{-0.48}^{+0.27}$& $\mathcal{U}[-10, 10]$ \\
$\sqrt{v\sin i_{\star}}\sin\lambda$ &$-2.21_{-1.02}^{+1.35}$ & $-1.66_{-0.87}^{+1.32}$ & $-1.56_{-0.79}^{+0.91}$ & $\mathcal{U}[-10, 10]$ \\
$\gamma$ ($\rm{m}\,\rm{s}^{-1}$) & $-1.89_{-14.18}^{+16.61}$& $-9.85_{10.40}^{+13.70}$ & $-8.77_{-8.74}^{+11.71}$& $\mathcal{U}[-1000, 1000]$ \\
$\dot{\gamma}$ ($\rm{m}\,\rm{s}^{-1}\,\rm{d}^{-1}$) & $167.7_{-140.9}^{+116.5}$& $205.59_{-132.07}^{+98.18}$ & $170.19_{-121.64}^{+101.77}$ & $\mathcal{U}[-1000, 1000]$ \\
$\ddot{\gamma}$ ($\rm{m}\,\rm{s}^{-1}\,\rm{d}^{-2}$)  & $-324.0_{-413.9}^{+518.4}$& $-294.815_{-326.17}^{+419.60}$ & $-154.79_{-362.68}^{+400.28}$ & $\mathcal{U}[-1000, 1000]$ \\
$\ln \sigma$ ($\rm{m}\,\rm{s}^{-1}$)  & $0.86_{-0.56}^{+0.50}$ & $0.88_{-0.57}^{+0.51}$ & $0.91_{-0.57}^{+0.49}$ & $\mathcal{U}[0, \ln(30)]$ \\
$u_{1,\rm{TESS}}$& 0.4006 & 0.4006 & 0.4006 & Fixed\\
$u_{2,\rm{TESS}}$& 0.2243 & 0.2243 & 0.2243 & Fixed \\
$u_{1,\rm{RM}}$& 0.6436 & 0.6436 & 0.6436 & Fixed \\
$u_{2,\rm{RM}}$& 0.1311 & 0.1311 & 0.1311 & Fixed \\
\hline
$\left|\lambda\right|$ ($^{\circ}$) & $33_{-20}^{+17}$& $25_{-20}^{+15}$ & $24_{-14}^{+14}$ & Derived \\
$v\sin i_{\star}$ ($\rm{km}\,\rm{s}^{-1}$) &$16.69_{-2.40}^{+2.72}$ & $15.21_{-0.88}^{+0.88}$ & $14.36_{-0.48}^{+0.49}$ & Derived$^{\ddagger}$ \\
$\rho$ ($\rho_{\odot}$) & $0.59_{-0.15}^{+0.18}$ & $0.65_{-0.15}^{+0.13}$ & $0.64_{-0.13}^{+0.14}$&  Derived$^{\ddagger}$\\
$R_{\rm p}$ ($R_\oplus$)  &$4.87_{-0.26}^{+0.26}$ & $4.84_{-0.25}^{+0.26}$ & $4.85_{-0.25}^{+0.26}$ & Derived\\
$T_{14}$ (hrs) & $4.69_{-0.39}^{+0.48}$&  $4.54_{-0.27}^{+0.43}$ &  $4.57_{-0.27}^{+0.41}$ & Derived \\  
$b$  & $0.45_{-0.24}^{+0.14}$& $0.39_{-0.22}^{+0.16}$ & $0.40_{-0.18}^{+0.15}$& Derived \\
$i_{\rm{orb}}$ ($^{\circ}$) & $88.43_{-0.69}^{+0.91}$ & $88.70_{-0.71}^{+0.78}$ & $88.63_{-0.66}^{+0.66}$& Derived\\ 
$a/R_\star$  & $16.5_{-1.5}^{+1.5}$& $17.1_{-1.5}^{+1.1}$ &  $17.0_{-1.4}^{+1.1}$& Derived  \\
$\sigma$ ($\rm{m}\,\rm{s}^{-1}$) &$2.37_{-1.02}^{+1.53}$  & $2.41_{-1.05}^{+1.61}$ & $2.50_{-1.08}^{+1.59}$ & Derived\\
\enddata
\tablecomments{Models 1, 2, and 3 use a fixed eccentricity of $e=0$ and share the same priors, as determined from initial MCMC runs, except for $v\sin i_{\star}$.  $^{\ddagger}$These two parameters are derived from RM effect.}
\end{deluxetable*}

TOI-942 was observed by TESS during its Sector 5 (30-minute cadence) and 32 (2-minute cadence). We obtained its light curve from the Mikulski Archive for Space Telescopes website\footnote{\url{https://archive.stsci.edu}} and detrended the stellar variation. To determine the sky-projected obliquity, we performed a combined analysis of the in-transit light curve and the RM effect. Prior to modeling, the TESS light curves were segmented into individual transit events to facilitate the analysis process.

For light curve modeling, we adopted the \texttt{Batman} package \citep{Kreidberg2015}. 
We used a quadratic law for limb-darkening effect, and the coefficients ($u_{1,\rm{TESS}}$, $u_{2,\rm{TESS}}$), 
were fixed to the results which were used in \citet{Wirth2021}. 
Since the previous three study on this system showed likely eccentric orbit for planet c ($e<0.8$ with 3$\sigma$ upper limit for \citealt{Zhou2021}, $e=0.175^{+0.139}_{-0.103}$ for \citealt{Carleo2021}, and $e=0.326^{+0.23}_{-0.16}$ for \citealt{Wirth2021}), the eccentricity was at first set to be non-zero and fitted with the longitude of periastron using the parameter set $\sqrt{e}\cos\omega$ and $\sqrt{e}\sin\omega$.
Other parameters for light curve modeling were set to be free parameters, including the logarithm of stellar density $\ln\rho_{\star}$, the orbital period $P_{\rm{orb}}$, the planet/star radius ratio $R_{\rm{p}}/R_{\star}$, the time of conjunction $t_0$ 
and cosine of orbital inclination $\cos i_{\rm{orb}}$ with uniform priors. The boundary of $\cos i_{\rm{orb}}$ was initially set between -1 and 1, and its value was then converted to the impact factor $b$. 

When modeling the RM effect, we followed the methodology described by \citet{Hirano2011}, which takes into account stellar rotation, macroturbulence, thermal broadening, pressure broadening, and instrumental broadening. The macroturbulence was calculated from the scaling relation given in \citet{Valenti2005}, thermal broadening and instrumental broadening were determined according to the formulation in \citet{Hirano2011}, and the pressure broadening was empirically set to be $1\,\rm{km}\,\rm{s}^{-1}$ for a K0 star \citep{Hirano2011}.
For limb-darkening effect, we also used a quadratic law with the coefficients ($u_{1,\rm{RM}}$, $u_{2,\rm{RM}}$) derived by \texttt{EXOFAST}\footnote{\url{https://astroutils.astronomy.ohio-state.edu/exofast/limbdark.shtml}} with input stellar parameters ($T_{\rm{eff}}$, $\rm{[Fe/H]}$, $\log g$) obtained from Section \ref{sec:star}. 
The free parameters includes sky-projected obliquity ($\lambda$) and projected rotational velocity $v\sin i_{\star}$. To improve sampling efficiency, they are transformed to ($\sqrt{v\sin i_{\star}}\cos\lambda$ and $\sqrt{v\sin i_{\star}}\sin\lambda$). 
In addition, RV offset $\gamma$, linear RV trend $\dot{\gamma}$, and quadratic RV trend $\ddot{\gamma}$ were set to model the intra-night RV drift caused by stellar activity and planets' orbital motion. An extra jitter term in logarithm $\ln\sigma_{\rm{jit}}$ is also included to account for any kinds of additional noise. Uniform priors are set to these parameters. 

We sampled the combined likelihood of the RM and transit model with a Markov Chain Monte Carlo (MCMC) algorithm, imlemented in the \texttt{emcee} \citep{Foreman-Mackey2013} code. We generated 64 random walkers, and their initials' positions were perturbed best-fit models with fixed $\sigma_{\rm{jit}}=0$. The best-fits were calculated by \texttt{Levenberg-Marquardt} method implemented in \texttt{lmfit} \citep{Newville2014}. 

We first generated several test runs, specifically with a non-zero uniform prior on eccentricity and a uniform prior on $v\sin i_{\star}$. Consequently and by-eye, we found that the maximum posterior of $\left| \lambda \right|$ was approximately $30^{\circ}$. We also noticed that $\sqrt{e}\cos\omega$ and $\sqrt{e}\sin\omega$ were poorly constrained and the maximum posterior of $e$ (converted) was at $0$ with $e<0.24$ at 84\%. Therefore, we defined the planet transiting in the $\cos\,i_{\rm{orb}}>0$ side and set $e=0$ by fixing $\sqrt{e}\cos\omega$ and $\sqrt{e}\sin\omega$ to zero, and generated long chains with narrower uniform priors to derive the final result. These new uniform priors are listed in Table \ref{tab:planet_para}.

To obtain the result, we stopped the chain once it passed the Gelman-Rubin test ($\rm{GR} < 1.03$). Consequently, we obtained a sky-projected obliquity of $\left|\lambda\right| = 33^{+17\circ}_{-20}$ from the posterior, favoring a prograde and likely aligned orbit for TOI-942c. We also obtained a projected stellar rotational velocity of $v\sin i_{\star} = 16.69_{-2.40}^{+2.72}\, \rm{km}\,\rm{s}^{-1}$, roughly consistent with the spectroscopic result $15.0\pm1.0\, \rm{km}\,\rm{s}^{-1}$ within 1$\sigma$ level. However, it is worth noting that the RM derived $v\sin i_{\star}$ was larger than any one from spectroscopy (Table \ref{tab:planet_para}). This suggested an overestimate on the stellar rotation, and so did the correlating parameter sky-projected obliquity in the fitting. The overestimate on $v\sin i_{\star}$ and $\left|\lambda\right|$ might be caused by the partial coverage of the transit. Therefore, in order to avoid the overestimate on stellar rotation, we generated MCMC chains with Gaussian priors on $v\sin i_{\star}$. We tested two Gaussian priors following the spectroscopic results. One is $\mathcal{N}[15.0, 1.0]$ based on this work, and another one is $\mathcal{N}[14.26, 0.50]$, a same prior as in \citet{Wirth2021}. As a result, we obtained $\left|\lambda\right| = 25^{+15\circ}_{-20}$ using the our spectroscopic results as prior and $\left|\lambda\right| = 24^{+14\circ}_{-14}$ from \citet{Wirth2021}'s prior. Free parameters and derived parameters with their uncertainties can be found in Table \ref{tab:planet_para}. By comparing different constraints, it is clear that with proper prior to the $v\sin i_{\star}$, the overestimate on $\left|\lambda\right|$ was reduced to some extent.

\section{Discussion}\label{sec:discussion}
\subsection{True Obliquity}
\begin{figure*}[ht]
\centering
\includegraphics[width=0.95\linewidth]{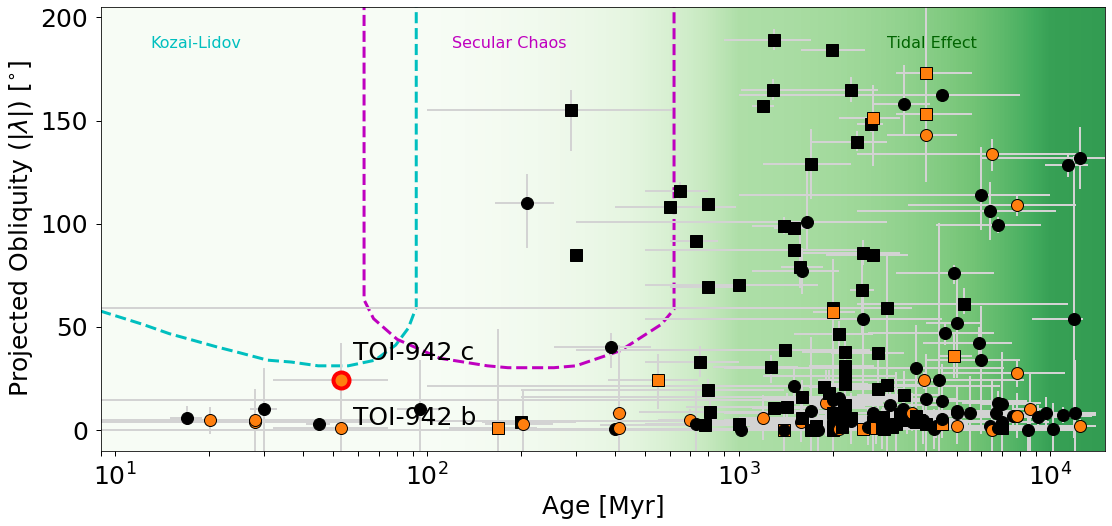}
\caption{The absolute values of projected stellar obliquity $\left| \lambda \right |$ against host star age. Planets in single-planet systems are marked in black while those in multiplanet systems are marked in red. Host stars having effective temperature lower and higher than 6100 K (Kraft break; \citealt{Kraft1967}) are respectively marked by circles and squares. More specifically, TOI-942c is highlighted by red edge. Different obliquity-exciting mechanisms on different timescales are marked by dashed lines in different colors, e.g., Kozai-Lidov mechanism, with a timescale shorter than a hundred Myr, and secular interaction, with a timescale in the magnitude of hundred Myr. Tidal realignment, which operates after a few Hundred Myr, is shaded green with darker color indicating a stronger effect. \label{fig:proj_obl_dist}}
\end{figure*}

TOI-942 possesses a well-determined stellar rotation period of $P_{\rm{rot}}=3.40 \pm 0.37$ days \citep{Zhou2021} or $P_{\rm{rot}}=3.39 \pm 0.01$ days \citep{Carleo2021}. We adopt a tighter constrained result of $P_{\rm{rot}}=3.39 \pm 0.01$ days. By combining this rotation period with the projected rotational velocity $v\sin i_{\rm{\star}}$, we can derive the posterior probability distribution for the true obliquity employing the methodology outlined in \citet{Masuda2020}. According to the equation proposed by \citet{Fabrycky2009}, the true obliquity $\psi$ can be expressed as:
\begin{equation}
    \cos \psi = \cos i_{\rm{orb}} \cos i_{\rm{\star}} + \sin i_{\rm{orb}} \sin i_{\rm{\star}} \cos \lambda,
\end{equation}
where $i_{\rm{\star}}$ is the stellar inclination, $i_{\rm{orb}}$ is the orbital inclination, and $\lambda$ is the sky-projected obliquity. Given that Model 3 in Table \ref{tab:planet_para} provides the best constraint on the sky-projected obliquity with $\left|\lambda\right| = 24^{+14\circ}_{-14}$, we utilized this result along with $v\sin i_{\star} = 14.26\, \rm{km}\,\rm{s}^{-1}$ \citep{Wirth2021} to derive the true obliquity. Here we include a conservative uncertainty of $2\, \rm{km}\,\rm{s}^{-1}$ on $v\sin i_{\star}$, especially considering that macroturbulence is not well known\footnote{Notably, \citet{Wirth2021} derived macroturbulence with a relatively strong Gaussian prior of $4.1 \pm 0.5\, \rm{km}\,\rm{s}^{-1}$.}. Consequently, we obtained $\psi = 30^\circ$ for maximum posterior and $\psi < 43^\circ$ at 84\% confidence for a positive $\cos i_{\rm{orb}}$ case, indicating a prograde orbit. Additionally, we conducted another test using the sky-projected obliquity from Model 2 in Table \ref{tab:planet_para} and $v\sin i_{\star}=15.0\, \rm{km}\,\rm{s}^{-1}$ (from our HIRES spectroscopy) with a conservative uncertainty of $2\, \rm{km}\,\rm{s}^{-1}$. This test yielded $\psi < 46^\circ$ at 84\% confidence. 

However, the physical context must be carefully considered when discussing the true obliquity of TOI-942c. First, both TOI-942b and c transit the star with observed orbital inclinations close to $90^\circ$, suggesting that planets b and c are likely coplanar with a small mutual inclination\footnote{The mutual inclination $\Delta i_{\rm{orb}} = \left| i_{\rm{orb,1}} - i_{\rm{orb,2}} \right|$ calculated from transit modeling provides a lower limit} of $\Delta i_{\rm{orb}}=0.8^{\circ}$.

We also know that planets b and c are not in mean-motion resonance, meaning the two planets are not coupled. In such a case, a mutual inclination up to twice the true obliquity could be induced because the rapidly rotating star's oblateness can generate a quadrupole moment and torque the orbit of each planet \citep{Spalding2016}. If the true obliquity is as large as the estimated $\sim$30$^{\circ}$, the mutual inclination excited by the quadrupole moment from oblateness could vary in a sinusoidal pattern, with a maximum of twice the obliquity, i.e., $\sim$60$^{\circ}$. Given that TOI-942b orbits in a well-aligned manner \citep{Wirth2021} and has a small observed mutual inclination with planet c, it is more likely that planet c is also aligned and coplanar with planet b.

Further, we roughly estimated the rate of nodal precession assuming the large obliquity for TOI-942c. The precession rate can be expressed as:
\begin{equation}
    \frac{\mathrm{d}\Omega}{\mathrm{d}t} = \frac{\partial \mathcal{H}}{\partial H}
    \label{eqn:precession}
\end{equation}
where $\Omega$ is ascending node, $\mathcal{H}$ is the Hamiltonian of planetary orbital evolution, and $H$ is defined as $H = \cos i$, where $i$ is planetary inclination. The Hamiltonian $\mathcal{H}$ includes two types of interactions: the quadrupole interaction generated by stellar oblateness and the secular interaction between planets. We assumed circular orbit for simplicity and utilized the equation (3) in \citet{Spalding2016} in our calculation:
\begin{align}
\mathcal{H}=\underbrace{\frac{G m m^{\prime}}{a^{\prime}}\bigg[\big(s^2+s^{\prime2}\big)f_3+ss^{\prime}f_{14}\cos(\Omega-\Omega^{\prime})\bigg]}_{\textrm{Planet-planet: Secular interaction}}\nonumber \\
\underbrace{-\frac{3G m M_\star}{a}J_{2}\bigg(\frac{R_\star}{a}\bigg)^2s^2-\frac{3G m^{\prime} M_\star}{a^{\prime}}J_{2}\bigg(\frac{R_\star}{a^{\prime}}\bigg)^2s^{\prime2} }_{\textrm{Planet-star: quadrupole interaction}},\label{eqn:Hamilton}
\end{align}
where the prime superscript is to distinguish two planets in the system. In the Hamiltonian, $s$ is defined as $\sin(i/2)$, and factors $f_3$ and $f_{14}$ are:
\begin{equation}
    f_3 = -\frac{1}{2}f_{14} = -\frac{1}{2}\left(\frac{a}{a^{\prime}}\right)b_{(1)}^{3/2}\left(\frac{a}{a^{\prime}}\right),
\end{equation}
where $b^{(1)}_{3/2}(\alpha)$ is the Laplace coefficient:
\begin{equation}
    b_{3/2}^{(1)}(\alpha)\equiv\frac{1}{\pi}\int^{2\pi}_0\bigg[\frac{\cos \psi}{(1+\alpha^2-2\alpha\cos \psi)^{3/2}}\bigg]d\psi.
\end{equation}
For other factors, $G$ is the gravitational constant, $m$ is the planetary mass, $M_{\star}$ is the stellar mass, $a$ is the semimajor axis, $R_{\star}$ is the stellar radius, and $J_{2}$ is the stellar oblateness. The estimate of $J_{2}$ can be found in Section 3.1 of \citet{Spalding2016}. It is also worth mentioning that, in a two-planet system, both planets are expected to have an equal precession rate due to secular interaction (e.g., \citealt{Bailey2020}).

For TOI-942, we simply assumed both planets have mass of $10\,M_{\oplus}$ (arbitrarily assumed), planet b is initially well aligned \citep{Wirth2021}, and planet c has a true obliquity of $30^{\circ}$. That is to say, we set an initial mutual inclination of $\Delta i_{\rm{orb}}=30^{\circ}$ regardless of its formation scenario. We estimated an oblateness of $J_{2} = 7.7 \times 10^{-5}$ according to the \citet{Spalding2016}. We separately calculated the the nodal precession rate induced by oblateness and secular interactions for TOI-942. Consequently, we estimated the nodal precession rate to be at a magnitude of $\sim$0.01$^\circ$/yr due to oblateness and $\sim$1$^\circ$/yr due to secular interactions for both planets.  Notably, such a precession can result in hour-long transit duration variation within 2 years for TOI-942. However, TESS data did not show any transit duration variation for both TOI-942b and c, strongly disfavoring the high-obliquity assumption for TOI-942c. 

In general, if two planets in a compact system both transit the host star, their mutual inclination must be small to ensure that their orbits consistently intersect the stellar surface from the observer's viewpoint. Otherwise, due to the high nodal precession rate, transit duration variations would be detected, and double transit would be inconsistent. For TOI-942 system, the largest mutual inclination can be theoretically $\Delta i_{\rm{orb, max}}=i_{\rm{b,max}}+i_{\rm{c,max}} \approx R_{\star}/a_{\rm{b}} + R_{\star}/a_{\rm{c}} = 8^{\circ}$ to maintain consistent transit for both planets. Given the mutual inclination can be doubled from the quadrupole moment generated by the stellar oblateness, the maximum true obliquity of both planets in the TOI-942 system should theoretically not exceed $4^{\circ}$. To sum up, the dynamic analysis strongly contradicts the best-fit obliquity obtained from RM modeling and, it instead favors a well-aligned and co-planar scenario, which is about 2.3-$\sigma$ biased from the best-fit. This suggests that the sky-projected and true stellar obliquity result of TOI-942c should be overestimated probably owing to the incomplete coverage of the transit event.

\subsection{Obliquity in Time}
Figure \ref{fig:proj_obl_dist} displays all the planets having the projected stellar obliquity and stellar age estimates\footnote{Data acquisition from TEPCat \url{https://www.astro.keele.ac.uk/jkt/tepcat/obliquity.html} \citep{Southworth2011} and NASA Exoplanet Archive \url{https://exoplanetarchive.ipac.caltech.edu/}}. As more stellar obliquities of young planetary systems have been measured, two trends between stellar obliquity and stellar age emerge: (\textit{a}) systems younger than 100 Myr are predominantly well-aligned, and (\textit{b}) there are no strongly-misaligned multiplanet systems younger than 1 Gyr. 
The trend (\textit{a}) supports that orbit-tilting mechanisms which only operate during the disk-hosting stage (e.g., \citealt{Lai2011, Batygin2012}) is likely not the dominant channel to generate spin-orbit misalignment \citep{Dai2020, Johnson2022}. 
The trend (\textit{b}) strongly indicates that planets in young multiplanet systems, like the 50-Myr-old TOI-942 system, migrated inward smoothly within the disk, underwent dynamically cool formation, and did not experience re-alignment prior to 1 Gyr. It is noteworthy that misalignment may occur in the early stages; for instance, K2-290 harbors two coplanar retrograde planets, believed to form through disk tilting by its close stellar companion K2-290 B \citep{Hjorth2021} during the disk-hosting stage or via gravitational perturbations from its outer stellar companion K2-290 C \citep{Best2022} over a longer timescale (exceeding 100 Myr). However, due to the limited number of obliquity estimates for young multiplanet systems, none have been observed to undergo misalignment such as disk tilting. Building up a larger sample of young multiple planetary systems with stellar obliquity measurement holds the promise of constraining the origin of the misalignment. 

\subsection{Laboratory for Atmospheric Loss}

\begin{figure*}[ht]
\centering
\includegraphics[width=0.85\linewidth]{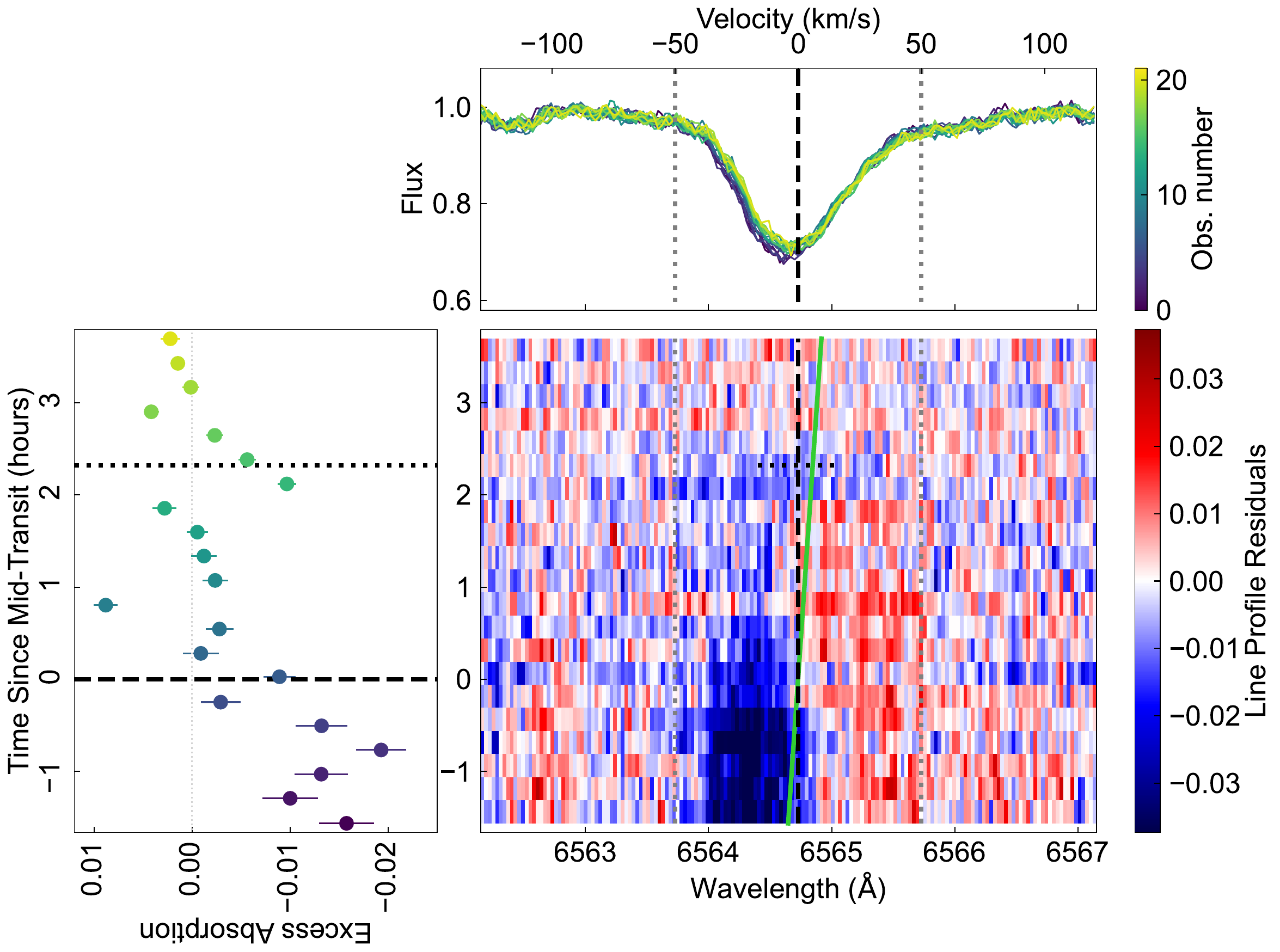}
\caption{Top: Normalized spectra of H$\alpha$ line \AA\, during transit event of planet c with their color indicating their observation number (epoch). The dashed line marks the line center at 6564.7 \AA, and the dotted lines indicate 1 \AA\, from the line center. Left: Excess absorption of H$\alpha$ variation by time, with the absorption referenced by an average post-egress spectrum, and calculated in a bandpass of 2 \AA\, centered at 6564.7 \AA. The color scale is the same for the top sub-plot. The dashed line and dotted line indicate mid-transit and egress, respectively. Notably, ingress is not included. Main (Bottom right): The measured H$\alpha$ line profile residuals as a function of time and wavelength. The dashed line and vertical dotted line correspond to the top sub-plot the top subplot, while the horizontal dotted line indicates the egress as in the left subplot. The predicted planetary velocity is marked by green solid line. An excess absorption and blueshift can be seen in dark blue. \label{fig:halpha}}
\end{figure*}

Atmospheric loss via photoevaporation caused by high-energy radiation from the host star \citep{Owen2017, Ginzburg2018} is thought to be the formation of hot Neptune desert as well as the the bimodal radius distribution of sub-Neptunian planets \citep{Fulton2017}. Also, \citet{Fulton2017} showed these small planets from \textit{Kepler} are normally smaller than $3.5\,R_{\oplus}$. TOI-942b and c has radii of 3.89 $R_\oplus$ \citep{Wirth2021} and 4.76 $R_\oplus$, respectively, which are apparently larger than sub-Neptunes from \textit{Kepler}. It suggests the TOI-942b and c may still be still contracting from the heat of formation, and as such are suitable for a comparative study because the planets are bathed in the same high-energy radiation environment except for a difference in orbital distance.

Our spectroscopic observation covers the wavelength region around 6564 \AA, enabling an examination of the H$\alpha$ line to search for excess absorption. As illustrated in Figure \ref{fig:halpha}, a blueshift accompanied by a deeper absorption was observed before the mid-transit and seemed to disappear after that. Three hypotheses could account for this phenomenon.

One plausible explanation is the extended planetary atmosphere. The blueshifted absorbing H$\alpha$ suggests an acceleration of the planetary atmosphere away from the star (e.g., \citealt{Benatti2021,Feinstein2021,Lissen2023,Yan2024,Orell-Miquel2024}). If the excess absorption starts from the ingress and continues after egress, it could signify a "comet-like" tail of the planetary atmosphere trailing the planet, absorbing stellar photons while accelerated away by stellar winds \citep{Tyler2024}. Unfortunately, the absence of pre-ingress data disabled the determination of the absorption depth prior to transit. Following up observations, covering from pre-ingress to (at least) in-transit may address this limitation. Furthermore, if atmospheric loss is indeed the cause, we expect to observe a similar blueshifted excess absorption manifesting in the {He\,{\footnotesize I}} triplet at 10833 \AA during the full transit event (e.g., \citealt{Zhang2022, Tyler2024}).

Two alternative hypotheses include instrumental variability and stellar activity. While slit instability in the Keck/HIRES optical path could induce spectral line fluctuations, such variation would extend across the entire spectrum rather than being confined to a specific line. Given the absence of comparable spectral variations across the entire spectrum, e.g., {Na\,{\footnotesize I}}, {Ca\,{\footnotesize I}}, {Ca\,{\footnotesize II}}, and {Cr\,{\footnotesize I}} lines \citep{Lissen2023}, the instrumental explanation is unlikely. Stellar activity is another scenario. Crossing spots on the stellar surface could yield similar spectral line variations (e.g., \citealt{Feinstein2021, Wirth2021}), yet this fails to explain the absence of concurrent variation across the entire spectrum. Since H$\alpha$ is a sensitive indicator for chromospheric activity,  this possibility warrants careful consideration. We assessed the $S$-index of {Ca {\footnotesize II}} H\&K lines \citep{Isaacson2010} in our spectra but found no strong correlation ($r=0.19$) between the H$\alpha$ excess absorption and the $S$-index. This suggests that chromospheric activity alone may not account for the observed blueshifted excess absorption.

In summary, while this study cannot confirm the true cause of the blueshifted excess absorption, atmospheric loss emerges as a promising hypothesis. Future observations hold the potential to tell us the truth.

H.Y.T appreciates the support by the EACOA/EAO Fellowship Program under the umbrella of the East Asia Core Observatories Association. This research is supported by the National Natural Science Foundation of China (NSFC) under grant No.12073044 and  U2031144. 
This research has made use of the NASA Exoplanet Archive, which is operated by the California Institute of Technology, under contract with the National Aeronautics and Space Administration under the Exoplanet Exploration Program. 
We thank the time assignment committees of the University of California, the California Institute of Technology, NASA, and the University of Hawai'i for supporting the TESS-Keck Survey with observing time at the W. M. Keck Observatory.
We gratefully acknowledge the efforts and dedication of the Keck Observatory staff for support of HIRES and remote observing. We recognize and acknowledge the cultural role and reverence that the summit of Maunakea has within the indigenous Hawaiian community. We are deeply grateful to have the opportunity to conduct observations from this mountain.

\facility{Keck I (HIRES), TESS, Exoplanet Archive}
\software{Batman \citep{Kreidberg2015}, Emcee \citep{Foreman-Mackey2013}, EXOFAST \citep{Eastman2013}, Isoclassify \citep{Huber2017}, lmfit \citep{Newville2014}, SpecMatch \citep{Petigura2015, Yee2017}}

\appendix
\section{Additional Data and Figure}
\begin{figure*}[ht]
\centering
\includegraphics[width=0.85\linewidth]{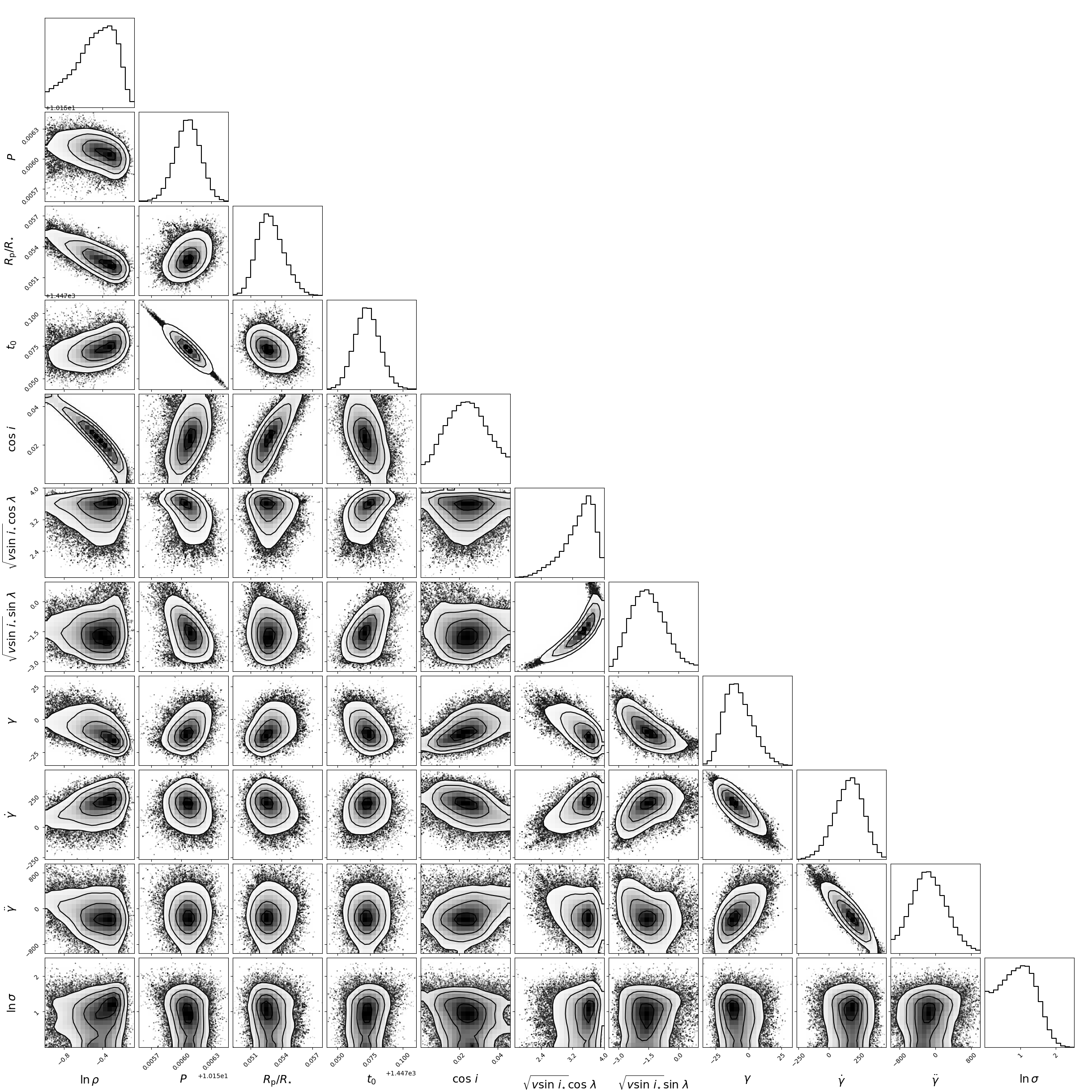}
\caption{The posterior distribution obtained from the MCMC chain of Model 3 in Table \ref{tab:planet_para} (i.e., the best model). \label{fig:corner}}
\end{figure*}

The Keck/HIRES data obtained during the transit are listed in Table \ref{tab:data}, and the posterior distribution of Model 3 in Table \ref{tab:planet_para} (i.e., the best model) is shown in Figure \ref{fig:corner}.

\begin{deluxetable*}{cccccccc}
\tablecaption{Keck/HIRES data during transit} \label{tab:data}
\tablehead{
BJD-2457000 & RV $(\rm{m}\, \rm{s}^{-1})$& $\sigma_{\rm{RV}}\, (\rm{m}\,\rm{s}^{-1})$ & $S_{\rm{HK}}$ & $\sigma_{S_{\rm{HK}}}$ & $\rm{EW}_{\rm{H}\alpha}$ & $\sigma_{\rm{EW}_{\rm{H}\alpha}}$ & Airmass}
\startdata
\hline
2543.862651 &  20.22 & 5.20 & 0.736 &    0.001 & -0.0157 & 0.0028 &     1.81 \\
2543.873960 &  12.04 & 4.22 & 0.738 &    0.001 & -0.0100 & 0.0028 &     1.69 \\
2543.884875 &   5.29 & 3.89 & 0.735 &    0.001 & -0.0132 & 0.0027 &     1.60 \\
2543.895789 & -14.18 & 4.01 & 0.736 &    0.001 & -0.0193 & 0.0026 &     1.52 \\
2543.906715 &   6.34 & 4.03 & 0.741 &    0.001 & -0.0132 & 0.0026 &     1.46 \\
2543.917398 &  -2.00 & 3.82 & 0.762 &    0.001 & -0.0029 & 0.0020 &     1.41 \\
2543.928891 &  -5.44 & 4.49 & 0.759 &    0.001 & -0.0089 & 0.0016 &     1.38 \\
2543.939446 & -19.19 & 4.02 & 0.752 &    0.001 & -0.0009 & 0.0018 &     1.35 \\
2543.950430 & -19.42 & 3.54 & 0.745 &    0.001 & -0.0028 & 0.0014 &     1.32 \\
2543.961230 & -27.44 & 4.47 & 0.750 &    0.001 &  0.0088 & 0.0012 &     1.31 \\
2543.972422 & -20.67 & 4.35 & 0.741 &    0.001 & -0.0024 & 0.0013 &     1.31 \\
2543.983429 & -34.43 & 4.69 & 0.741 &    0.001 & -0.0012 & 0.0013 &     1.31 \\
2543.994228 & -15.14 & 4.32 & 0.730 &    0.001 & -0.0005 & 0.0011 &     1.31 \\
2544.005015 &  -6.66 & 3.88 & 0.737 &    0.001 &  0.0028 & 0.0012 &     1.33 \\
2544.016022 &   3.64 & 3.70 & 0.736 &    0.001 & -0.0097 & 0.0009 &     1.35 \\
2544.026994 &  19.81 & 4.06 & 0.734 &    0.001 & -0.0056 & 0.0009 &     1.39 \\
2544.037966 &  18.29 & 3.87 & 0.741 &    0.001 & -0.0023 & 0.0009 &     1.43 \\
2544.048615 &  18.39 & 3.92 & 0.739 &    0.001 &  0.0042 & 0.0008 &     1.48 \\
2544.059703 &  26.08 & 4.27 & 0.743 &    0.001 &  0.0001 & 0.0009 &     1.55 \\
2544.070513 &  17.48 & 3.84 & 0.739 &    0.001 &  0.0014 & 0.0007 &     1.63 \\
2544.081624 &  17.10 & 4.10 & 0.735 &    0.001 &  0.0022 & 0.0010 &     1.73 \\
\hline
\enddata
\tablecomments{The raw echelle spectra are accessible from Keck data archive (\url{https://koa.ipac.caltech.edu/cgi-bin/KOA/nph-KOAlogin}). The weather condition at Maunakea during the observation was good, with mostly clear sky, an average humidity of 10\% both inside and outside the dome, and an average seeing of approximately 0.8 arcseconds, ranging between 0.4 and 1.6 arcseconds. More information about the observing weather condition can be found at \url{http://mkwc.ifa.hawaii.edu/}. }
\end{deluxetable*}

\bibliography{sample631}{}
\bibliographystyle{aasjournal}

\end{CJK*}
\end{document}